\documentclass[aps,prc,superscriptaddress,showpacs,secnumroman,showkeys]{revtex4}
\usepackage{amsmath,bm}%
\usepackage{mathrsfs}
\usepackage{graphics}
\usepackage{graphicx}
\usepackage{color}%
\usepackage{float}
\usepackage{color}

\usepackage[normalem]{ulem}  % \sout{old text} for strikeout

\renewcommand\sout{\bgroup \color{red} \ULdepth=-.5ex \ULset}
\begin{document}

\def\Journal#1#2#3#4{{#1} {{#2}}, {#3} (#4).}
\def\ANP{Adv. Nucl. Phys.}
\def\ARNPS{Ann. Rev. Nucl. Part. Sci.}
\def\CTP{Commun. Theor. Phys.}
\def\EPJA{Eur. Phys. J. A}
\def\EPJC{Eur. Phys. J. C}
\def\IJMPA{International Journal of Modern Physics A}
\def\IJMPE{International Journal of Modern Physics E}
\def\JCHP{J. Chem. Phys.}
\def\JCP{Journal of Computational Physics}
\def\JHEP{JHEP}
\def\JPCS{Journal of Physics: Conference Series}
\def\JPG{J. Phys. G: Nucl. Part. Phys.}
\def\NATURE{Nature}
\def\NC{La Rivista del Nuovo Cimento}
\def\NCA{IL Nuovo Cimento A}
\def\NPA{Nucl. Phys. A}
\def\NST{Nuclear Science and Techniques}
\def\PA{Physica A}
\def\PAN{Physics of Atomic Nuclei}
\def\PHY{Physics}
\def\PRA{Phys. Rev. A}
\def\PRC{Phys. Rev. C}
\def\PRD{Phys. Rev. D}
\def\PLA{Phys. Lett. A}
\def\PLB{Phys. Lett. B}
\def\PLD{Phys. Lett. D}
\def\PRL{Phys. Rev. Lett.}
\def\PL{Phys. Lett.}
\def\PREV{Phys. Rev.}
\def\PREP{\em Physics Reports}
\def\PROG{Progress in Particle and Nuclear Physics}
\def\RPP{Rep. Prog. Phys.}
\def\RDNC{Rivista del Nuovo Cimento}
\def\RMP{Rev. Mod. Phys}
\def\SCIENCE{Science}
\def\ZPA{Z. Phys. A.}

\def\ANN{Ann. Rev. Nucl. Part. Sci.}
\def\ANNAST{Ann. Rev. Astron. Astrophys.}
\def\AP{Ann. Phys}
\def\APJ{Astrophysical Journal}
\def\APJS{Astrophys. J. Suppl. Ser.}
\def\EJP{Eur. J. Phys.}
\def\LANC{Lettere Al Nuovo Cimento}
\def\NCA{Nuovo Cimento A}
\def\PHYS{Physica}
\def\NP{Nucl. Phys}
\def\MATH{J. Math. Phys.}
\def\JPAM{J. Phys. A: Math. Gen.}
\def\PRO{Prog. Theor. Phys.}
\def\NPB{Nucl. Phys. B}

%\begin{CJK*}{GBK}{}
\title{Decay Modes of the Hoyle State in $^{12}C$}
\author{H. Zheng}
%\email[]{zheng@lns.infn.it}
\affiliation{Laboratori Nazionali del Sud, INFN, via S. Sofia 62, I-95123 Catania, Italy;}
\author{A. Bonasera}
%\email[]{abonasera@comp.tamu.edu}
\affiliation{Laboratori Nazionali del Sud, INFN, via S. Sofia 62, I-95123 Catania, Italy;}
\affiliation{Cyclotron Institute, Texas A$\&$M University, College Station, Texas 77843, USA;}
\author{M. Huang}
%\email[]{huangmeirong05@163.com}
\affiliation{College of Physics and Electronics information, Inner Mongolia University for Nationalities, Tongliao, 028000, China.}
\author{S. Zhang}
%\email[]{zsylt0416@163.com}
\affiliation{College of Physics and Electronics information, Inner Mongolia University for Nationalities, Tongliao, 028000, China.}

%\date{\today}

\begin{abstract}
Recent experimental results give an upper limit less than 0.043\% (95\% C.L.) to the direct decay of the Hoyle state into 3$\alpha$ respect to the sequential decay into $^8${Be}+$\alpha$. We performed one and two-dimensional tunneling calculations to estimate such a ratio and found it to be more than one order of magnitude smaller than experiment depending on the range of the nuclear force. This is within high statistics experimental capabilities. Our results can also be tested by measuring the decay modes of high excitation energy states of $^{12}$C where the ratio of direct to sequential decay might reach 10\% at $E^*$($^{12}$C)=10.3 MeV. The link between a Bose Einstein Condensate (BEC) and the direct decay of the Hoyle state is also addressed. We discuss a hypothetical `Efimov state' at $E^*$($^{12}$C)=7.458 MeV, which would mainly {\it sequentially} decay with 3$\alpha$ of {\it equal energies}: a counterintuitive result of tunneling. Such a state, if it would exist, is at least 8 orders of magnitude less probable than the Hoyle's, thus below the sensitivity of recent and past experiments.
\end{abstract}

%\pacs{21.10.Tg, 23.60.+e, 27.20.+n, 03.65.Xp }


\maketitle
%\end{CJK*}

The changing shape of a nucleus depending on its excitation energy plays a very important role in nuclear reactions, such as the formation of $^{12}$C, $^{16}$O etc., essential for understanding the interiors of stars \cite{Ichikawa:2011iz, Marin-Lambarri:2014zxa, Epelbaum:2011md, Epelbaum:2013paa}. The linear chain structures constitute a longstanding challenge both in theoretical and experimental studies since Morinaga suggested this exotic arrangement in 1950s \cite{Morinaga:1956zza, Morinaga:1966pl}. Anti-symmetrized molecular dynamics (AMD) calculations, which allow the coexistence of cluster nuclei and shell-model-like aspects, predict a linear-like chain of $\alpha$ clusters, in which the largest angle of the vertices is more than 120 degree, for the $0^+_3$, $1^-_1$, $2^+_2$ and $2^-_1$ at 10.3, 10.844, 11.16 and 11.828 MeV respectively. In these states, one $\alpha$ cluster seems almost to be escaping. While the 3$\alpha$ clusters in the $0^+_2$, $3^-_1$ and $4^-_1$ states at 7.654, 9.641 and 19.55 MeV form isosceles triangle configurations close to equilateral triangles \cite{Enyo:2007zz}. However, there have been no experimental confirmations for 3$\alpha$ chain in $^{12}$C for those states. Recently, by studying the reaction $^{40}$Ca +$^{12}$C at 25 MeV/nucleon, it has been suggested that the Hoyle state has a minor decay branch $7.5\pm4.0$\% that produces three particles of almost equal energy and that such a decay provides evidence for Bose Einstein Condensate (BEC) \cite{Raduta:2011yz}. Later, by studying different projectile-target and beam energy combinations, a much lower limit was set for the component with three nearly equal-energy $\alpha$ particles \cite{Manfredi:2012zz, DellAquila:2017ppe, Smith:2017jub, Rana:2013hka}. The two most recent experiments at the time of writing \cite{DellAquila:2017ppe, Smith:2017jub, Kirsebom:view2017} give an upper limit for the ratio of the direct decay (DD) to sequential decay (SD) less than 0.043\% (=1/2500, i.e., 1 out of 2500 events). The main reason for the difficulty to determine different decay modes of the Hoyle state is due to the experimental acceptance. In fact modern silicon or other detectors type have a resolution in energy of the order of tens of keV. This number should be compared to the width of the Hoyle state, which is 8.5 eV \cite{c12expdata}. For instance, Raduta {\it et al.}  \cite{Raduta:2011yz} claim a $17.0\pm5.0$\% decay rate into the DD, but the {\it measured} width of their state is 330 keV due to the detector sensitivity. Increasing the detector sensitivity and the experimental technique results in a width of about 50 keV and corresponding decrease of the ratio to less than 0.5\% \cite{Kirsebom:2012zza} and further improved the upper limit to 0.2\% \cite{Itoh:2014mwa}, improving on the first result by Freer {\it et al.} \cite{Freer:1994zz}, less than 4\%. The latest improvements in Refs. \cite{DellAquila:2017ppe, Smith:2017jub, Kirsebom:view2017}, where higher statistics data was collected and further cuts were adopted by selecting events having an excitation energy narrowly distributed around the Hoyle state centroid at 7.65 MeV, gave values of 0.047\%  \cite{Smith:2017jub}  and 0.043\%  \cite{DellAquila:2017ppe}  respectively. Clearly, the more cuts/constraints are adopted the larger the statistics must be in order to obtain a more precise value for the DD. The lower limit quoted above corresponds to about 2500 events in the region of interest.

It is important to stress that the configurations found in microscopic calculations might not be recovered experimentally since the excited nuclei, in order to decay, have to tunnel the Coulomb barrier. Because of the tunneling, for instance for the Hoyle state, the condition where two $\alpha$ particles are close together to form a $^8$Be while the other $\alpha$ tunnels the barrier is more favorable. This is essentially the configuration (SD) found in experiments  \cite{Raduta:2011yz, Manfredi:2012zz, DellAquila:2017ppe, Smith:2017jub, Kirsebom:view2017, Kirsebom:2012zza, Freer:1994zz}. We will now discuss the conditions for which a configuration (say SD) is more favorable respect to another (say DD).  Let us first recall some known features of $^8$Be decay. This nucleus is short lived and decays into two alphas. The process can be understood through quantum mechanical tunneling.  The two $\alpha$-particles acquire about 91 keV after the decay. We can approximate the decay using the Gamow prescription, i.e., neglecting the nuclear attraction. This approximation is rather good when the final kinetic energies of the fragments are very small as compared to the Coulomb barrier. However, corrections due to the nuclear force can be easily implemented assuming a step function \cite{Kimura:2007rv}. The probability of tunneling at finite impact parameter, is given in terms of the action $A$ \cite{Bonasera:1994kdi, Bonasera:1997zz}:
\begin{equation}
\Pi[E-\frac{l(l+1)\hbar^2}{2I}] \sim e^{\frac{2}{\hbar}A}; A=\int_{R_N}^{R_0}dr \sqrt{2\mu(V_c(r)-E)}, \label{eq1}
\end{equation}
where $V_c(r)$, $R_N$ and $R_0$ are the Coulomb potential, the inner and the outer classical turning points, respectively.  $E$, $l$, $I$ and $\mu$ are the center of mass energy, the angular momentum, the moment of inertia and the reduced mass. In particular, if we consider the pure Coulomb penetrability, i.e., we take the limit of $R_N\rightarrow 0$, $l=0$, and no nuclear force, $A$ becomes $A_G=Z_T Z_P e^2 \pi \sqrt{\frac{\mu}{2E}}$: the Gamow limit. The lifetime is given by:
\begin{equation}
\tau = \tau_B(1+ e^{\frac{2}{\hbar}A}). \label{eq2}
\end{equation}
The $\Gamma_{1/2}$ width is given by $\Gamma_{1/2}=\frac{\hbar}{\tau \ln 2}$. The characteristic time for assault to the barrier $\tau_B$ is left as a free parameter and fitted to the lifetime of $^8$Be. Its value could be estimated from the average velocity of nucleons in the nucleus and its radius or other means, but since we are mainly concerned with ratios, $\tau_B$ cancels out. Also, to avoid a detailed description of the nuclear force, we assume that it overcomes the Coulomb repulsion for distances below $R_N=\alpha_N\times(R_1+R_2)$ and we vary the value of $\alpha_N$ from zero (Gamow limit) to 1 (touching spheres of radii $R_{1, 2}$). As we will show, the ratio is smoothly dependent on the value of $\alpha_N$ and future experimental results might confirm or disprove it. In order to reduce the variation of the ratio further, we can approximate the Coulomb potential to that of overlapping spheres when the relative distance between the separating nuclei is less than $R (=R_1+R_2)$
\begin{equation}
V_c(r)=\left\{\begin{array}{ll} 
\frac{Z_1Z_2e^2}{2R}(3-\frac{r^2}{R^2}), & \mbox{if  $r\le R$};\\
\frac{Z_1Z_2e^2}{R}, & \mbox{if $r\ge R$}.
\end{array}
\right.
\end{equation}
This term gives a modest modification of the results (at most a factor 2.5 for the Hoyle state  in the Gamow limit) thus giving more confidence to our approach and it will not be considered further. Eqs. (\ref{eq1}) and (\ref{eq2}) can be easily applied to the decay of $^8$Be and the SD of the Hoyle state. In the more general case of the Hoyle decay directly into 3$\alpha$, the action must be generalized in order to take into account the fact that tunneling is now occurring in the plane determined by the 3 particles, i.e., in two dimensions. We follow the spirit of Refs. \cite{Bonasera:1994kdi, Bonasera:1997zz} by defining collective variables {\bf P} and {\bf R}. In particular, in the case of equal energies of the 3 tunneling $\alpha$s the problem simplifies further since the relative momenta (distances) of the 3 particles are equal.  The action becomes $A = \int_{R_N}^{R_0}d{\bf R}{\bf P}$ which in two dimensions gives a factor of 2 after integrating over the relative angle between {\bf P} and {\bf R}, thus effectively reducing the two dimensional problem to a one dimensional as in Eq. (\ref{eq1}). Thus the action is calculated with the condition $r_{12}=r_{13}=r_{23}=r$ and similarly for the momenta. It is clear that different decay modes, for instance a linear chain, can be easily calculated with the conditions $r_{12}=r_{13}=r_{23}/2$, i.e., with particle 1 at the center of the chain. But these decay modes are also negligible respect to the SD. We expect a change less than a factor of 2 adding more configurations.  On the other hand, the ratios of DD/DDE for the Hoyle state are around 4 from different experimental measurements \cite{Rana:2013hka, Itoh:2014mwa}. DDE represents the three emitted $\alpha$s with similar energies in DD. Therefore, in this paper we will discuss the equal energy decays only. As we will see later, the ratio of DD/SD in our calculation is 40 times less than the experimental limit, our conclusion will not be affected by this simplification in our framework.

Under these assumptions it is straightforward to solve the action integral for different values of $\alpha_N$ and different excitation energies of $^{12}$C and angular momenta. In particular we have solved Eqs. (\ref{eq1}) and (\ref{eq2}) for the Hoyle, the 9.641 MeV ($J^\pi=3^-$) and the 10.3 MeV ($J^\pi=0^+$) states. As mentioned in the introduction, to these states we have added a hypothetical `Efimov state (ES)' \cite{Efimov:1970zz, Thomas:1935zz} at $E^*=7.458$ MeV, $J^\pi=0^+$. One of the reasons for discussing this state is on the question of the BEC explored by Raduta {\it et al.} \cite{Raduta:2011yz}. The authors claim that equal energies of the 3$\alpha$ might be a signature for a BEC.  We argue that such a claim could be not even true for non-interacting bosons since below the critical temperature we might have Bosons with zero energy and Bosons with energy larger than zero. We will show that for the ES the decay is mostly sequential, as for the Hoyle state, but the three $\alpha$s have the same final energy as for the DD! The fact that the 3$\alpha$ have the same energy gives a precise value of the excitation energy of the ES. Recall that the ES arises when no two body bound state exists but just a strong resonance while the three body system gets strongly bound \cite{fanophytoday76, Braaten:2006vd, greenephytoday10, efimov:nature09}. This result is based on the Thomas theorem discussed in the 30' \cite{Thomas:1935zz} and might apply well to $\alpha$ since $^8$Be is not bound while $^{12}$C is. Exploiting this fact one can define an ES when three particles are in mutual resonance with each other \cite{Braaten:2006vd, greenephytoday10, Tumino:2015jaa}, the vice versa might not be true \cite{Tumino:2015jaa}. This mechanism is similar to the exchange of a pion, which produces the nuclear force. The difference is that one of the 3 component particles is exchanged between the two others. In other words an $\alpha$ is exchanged between the other two and the relative kinetic energy of the 3$\alpha$ is exactly 91.84 keV($\times 3$), i.e., the gs of $^{8}$Be for each couple.  The exchange of the $\alpha$ particle results in a $1/R^2$ attractive potential only effective for $l=0$. This phenomenon has been demonstrated in atomic physics \cite{efimov:nature09, zaccanti:nature09} but not so far in nuclear physics for which it was first predicted. The main difficulty (apart Coulomb) is the short-range nuclear interaction and the scattering length, which are comparable. If we assume that the 3$\alpha$ are in mutual resonance to form $^{8}$Be, we can obtain the excitation energy of the state as: 
\begin{equation}
E^*(ES) = \frac{2}{3} \sum_{i=1, j>i}^{i=3}E_{ij}-Q =\frac{2}{3}(0.09184\times 3)+7.2747\mbox{ MeV} = 7.458 \mbox{ MeV}. \label{es3a}
\end{equation}
We would like to stress that we are not advocating here the existence of the ES but just show how tunneling could produce counterintuitive results. Detailed three-body calculations \cite{Suno:2016fjb, Ishikawa:2014mza} show a sharp resonance for the Hoyle state and a kink or bump depending on the used model in the region of the suggested ES. A possible interpretation of the kink is the transition between the nonresonant regime and the resonant regime \cite{Suno:2016fjb}. Of course, the possibility that the relative energy of the $\alpha$ satisfies exactly Eq. (\ref{es3a}) is rather {\it curios} and might call for further investigations on the reason why the Hoyle state is about 200 keV above such value \cite{Naidon:2016dpf}. The Hoyle energy is consistent with microscopic calculations of the Hoyle state, which do not give exactly equilateral triangles since only one $^8$Be can be formed and determines the dominant SD.  On the other hand, as we show below the probability of forming the ES is more than 8 orders of magnitude less than the Hoyle state \cite{Suno:2016fjb, Ishikawa:2014mza}, because of the lower available energy, and it could have been missed in the experimental investigations. For instance, in Refs. \cite{DellAquila:2017ppe, Smith:2017jub}, 28000 Hoyle state events were collected, way below what is needed to reveal the ES.
\begin{figure} [H]  
        \centering
       \begin{tabular}{c}
        \includegraphics[scale=0.4]{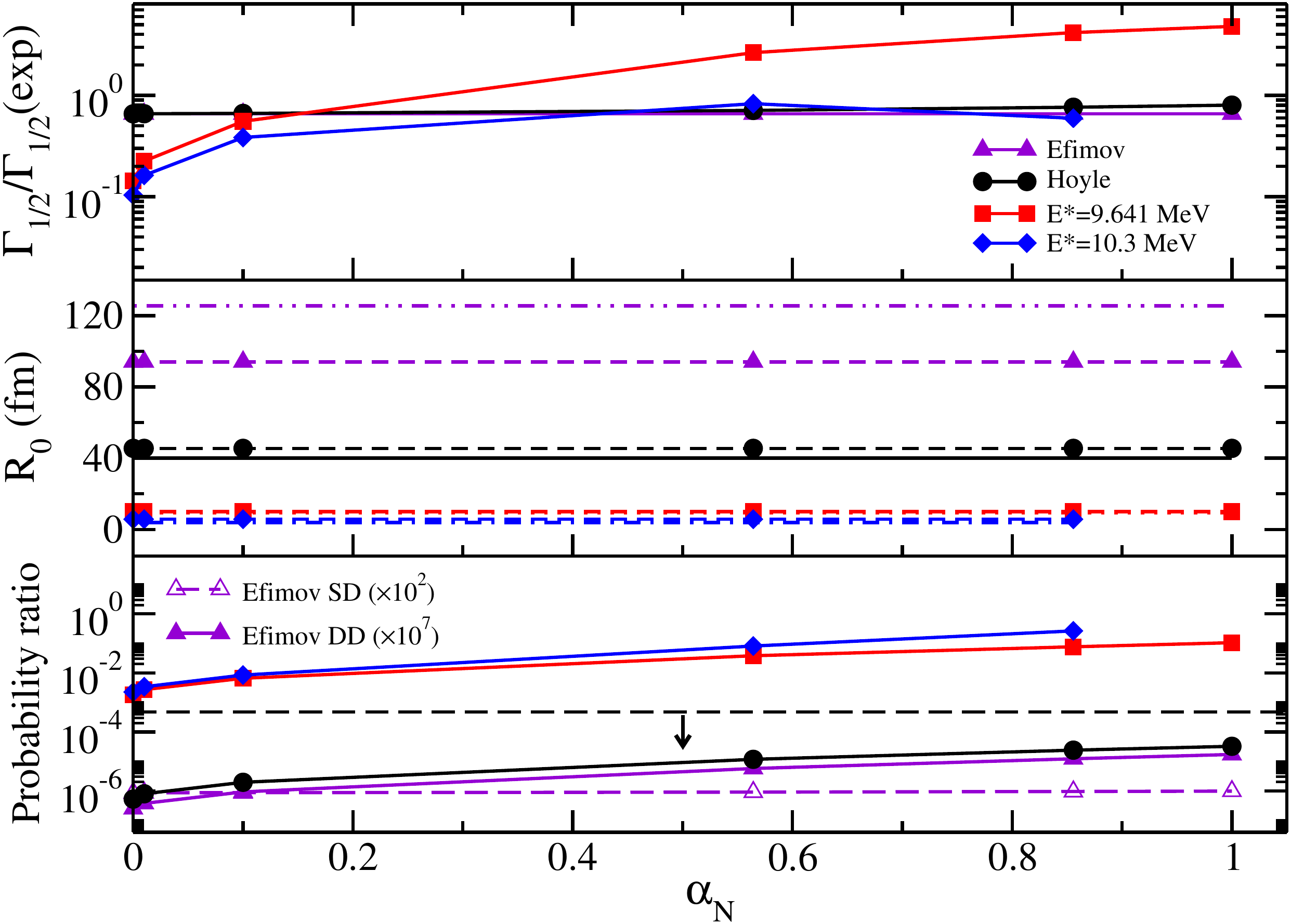}
\end{tabular}
\caption{(Color online) (Top) Theoretical widths normalized  by the experimental values, the hypothetical ES is divided by the experimental Hoyle state value. (Middle) Outer classical turning point for different excitation energies. Lines with symbols refer to the DD mechanism and the lines refer to the SD mechanism. (Bottom) The probability ratio of the DD to the SD mechanism. The dashed line with arrow gives the experimental upper limit value of the Hoyle state. The ES is divided by the Hoyle SD state both for the SD (dashed line with open triangles) and the DD (full line with full triangles).} \label{figure2}
    \end{figure}
    
 In Fig. {\ref{figure2} we plot some physical quantities obtained from the tunneling calculations as function of the parameter $\alpha_N$.  The top panel gives the ratio of the calculated widths to the experimental values. Since for the ES (full triangles) there are no experimental observations, we have divided the theoretical values by the Hoyle experimental result. Notice that all $E^*$ give a ratio approximately equal to 1 for $\alpha_N$ close to 0.6 which suggests that the nuclear force is indeed important to reproduce the data. The 9.641 MeV result (full squares) is slightly larger than 1 in the same region which might be due to the approximation in Eq. (\ref{eq1}) to take into account the angular momentum. Furthermore, since for the last case we can have 3 different momenta of inertia depending on the axis of rotation, we have considered the one giving largest probability, but averaging over the three different rotation axis does not change the results much. It should not surprise the fact that the ES has a width similar to the Hoyle state (full circles) since it is mainly due to the $^8$Be width and the dominance of the SD. The width in SD is defined as the width of the first decay $^{12}$C$\rightarrow ^{8}$Be+$\alpha$ plus the width of $^{8}$Be. Since the latter width is dominant it explains why we easily reproduce the experiment (recall that we have fitted the $\tau_B$ parameter in Eq. (\ref{eq2}) to reproduce the $^{8}$Be lifetime). In the middle panel, the outer classical turning point is reported. It shows that it largely decreases with increasing excitation energy of the state. From their values, we could derive the density for a system composed of $\alpha$ particles only (like in a star after burning the lighter elements) when they could possibly tunnel to make $^{8}$Be. As expected the ES would be more effective at the lowest densities followed by the Hoyle state. If we assume that $R_0$ plays the role of the scattering length we can derive if other ES would exist at smaller excitation energy. The number of states can be determined from the relation \cite{Efimov:1970zz}: $N=\frac{s_0}{\pi}\times \ln(R_0/r_0)\approx 1$, where the universal constant $s_0=1.0062378$, $r_0=\frac{\hbar}{m_\pi} = 1.4$ fm the range of the nuclear force ($m_\pi$ is the pion mass) and finally $R_0=\frac{1.44\times 2\times 2}{0.09184}=62.72$ fm from the g.s. decay of $^{8}$Be. The $R_0$ could be compared to the  neutron-proton scattering length which for keV energies is $l=\sqrt{\sigma/\pi} \approx 20$ fm, since the cross section is huge of the order of 20 b \cite{Hackenburg:2006qd}. Thus, if the ES would exist, it could be at the energy given above and consistent with the gs of $^8$Be. However recent three-body calculations \cite{Suno:2016fjb, Ishikawa:2014mza}, result in no explicit resonant state at energy below the Hoyle state energy, but curiously enough, they display a kink or a bump in this region, a further reason for more experimental and theoretical investigations keeping in mind that such small probabilities are a challenge both for detection and for extreme fine tuning of model parameters. We stress again that no experimental data support the ES and we are reporting as a {\it curious result} and that tunneling might give a SD but with 3 equal energies $\alpha$ particles as for DD. Of course, if such a state would exist, its decay mode could not be distinguished experimentally.  As we see from the middle panel, $R_0$ for the SD is much larger than the one for the DD. On the other hand for the other cases the $R_0$ is almost the same for SD and DD, with at most the DD slightly larger than SD. This is due to the decreased importance of the Coulomb barrier with increasing excitation energy.

In the bottom panel we plot the probability ratio of DD to SD as function of $\alpha_N$ for each excitation energy except for the ES for which we report the ratio of the SD (DD) divided by the SD of the Hoyle state to have a direct comparison of the two cases. The dashed line with arrow gives the experimental upper limit value for the ratio DD to SD of the Hoyle state. As we see, our calculated ratio is at least one order of magnitude smaller than the experimental upper limit, thus more precise experimental investigations might be able to fix its value. As we mentioned above, the ES is at least 8 orders of magnitude less probable than the Hoyle state.  Furthermore, the SD is dominant and at least 4 orders of magnitude larger than the DD. This is another {\it curious result} since it brings to mind the effect of three-body recombination where 3 bosons (atoms) collide to form a diatomic molecule and one atom \cite{Braaten:2006vd}, thus similar to our SD. 

\begin{figure} [H]  
        \centering
        \begin{tabular}{c}
        \includegraphics[scale=0.4]{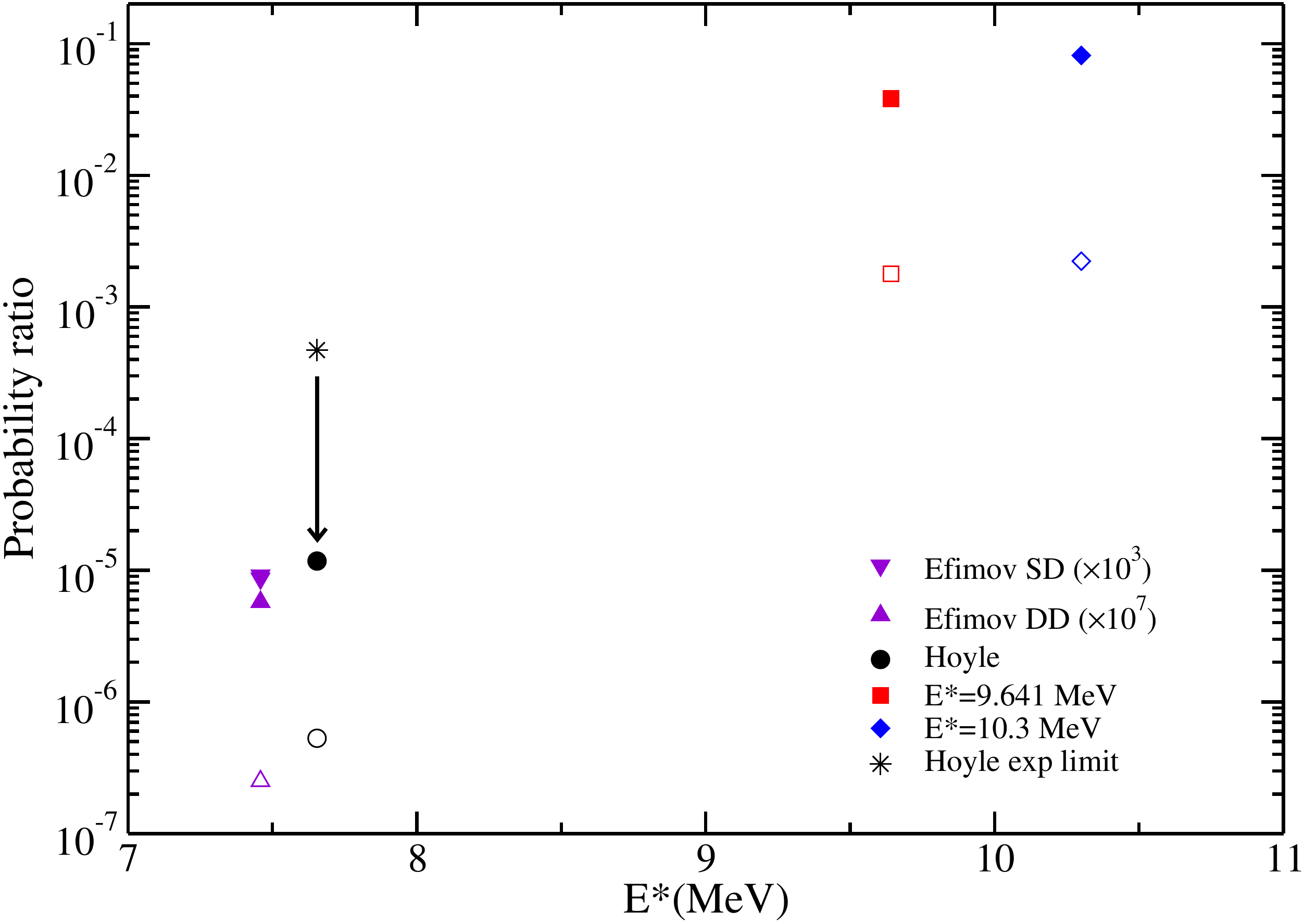}
\end{tabular}
    \caption{(Color online) Probability ratio as function of $^{12}$C excitation energy for fixed $\alpha_N$=0.56 (solid symbols) and $\alpha_N$=0 (open symbols). For the Efimov SD case, the solid down triangle and open down triangle cannot be distinguished since the results are so close for the two fixed $\alpha_N$. }\label{figure3}
    \end{figure}

The probability ratio is summarized in Fig. \ref{figure3} as function of the excitation energy and fixed $\alpha_N$=0.56 and $\alpha_N$=0. We show the results of Gamow  limit $\alpha_N=0$ for reference and will not discuss further. As expected the DD channel becomes more probable with increasing $E^*$ and it reaches 10\% at the highest value of $E^*$ considered here. The Hoyle state has a ratio more than a factor 10 smaller than the experimental upper limit and an experimental verification of this result would be interesting. Recall that the ES ratio is respect to the Hoyle state and as we see from Fig. \ref{figure3} the SD of the ES is more than 8 orders of magnitude smaller than the SD of the Hoyle state (solid down triangle). On the other hand the DD of the ES is about 12 orders of magnitude smaller respect to the SD of the Hoyle state. We would like to stress that this is the result of the tunneling calculation, other effects might be at play to modify this result.  For instance, the parameter $\tau_B$, Eq. (\ref{eq2}), might contain a contribution for two $\alpha$s to combine to make $^{8}$Be, or the angular momentum in the 9.641 MeV, $J^\pi=3^-$ case. For the ES, any combination of the 3$\alpha$ has the right energy to make $^8$Be. The real question at this point is not if the ES exists but why does not exist? In Ref. \cite{Tumino:2015jaa} it has been shown that in certain conditions 3$\alpha$ coming from higher excited states of $^{12}$C are emitted with relative energies which populate different states of $^{8}$Be at the same time, i.e., they are in mutual resonance. If there is no excited level in $^{12}$C at 7.458 MeV, we might still wonder what would happen in a $\alpha$-burning star if their average relative energies are close to the $^{8}$Be g.s.. Similarly in fragmentation reactions, in the later stages when the density is low and drops start to appear \cite{Schmidt:2016lpt, Mabiala:2016gpt, Marini:2015zwa} it is possible that this mechanism of mutual resonances increases the probability of making $^{12}$C even though some extra energy from the surrounding matter has to be provided to populate the Hoyle state (i.e., there is no ES).  The other peculiarity of the ES would be the fact that the final energies of the 3$\alpha$s are equal both for the SD and DD cases, thus they would be experimentally indistinguishable. It is worth to mention that the result of Ref. \cite{Ishikawa:2014mza} for the Hoyle state (i.e., the ratios of DDE/SD and DDL/SD are 0.005\% and 0.03\% respectively. DDL represents the three $\alpha$s are emitted collinear in DD) based on a completely different approach are in surprising agreement to ours (i.e., the ratio of DDE/SD is 0.0012\%. If we take into account that the three $\alpha$s are indistinguishable in DDE and only the two $\alpha$s in $^8$Be are indistinguishable in SD, the ratio of DDE/SD becomes 0.0036\% in our approach) suggesting that the physics entering the two approaches is the same and in particular the role of the nuclear force.

In conclusion, in this paper we have estimated the ratio of the direct decay to sequential decay for the Hoyle state and shown that it is at least an order of magnitude less than the current experimental upper limit.  We have demonstrated that such ratio depends essentially on the excitation energy because of the Coulomb barrier. We have postulated the existence of an `Efimov state' at $E^*$=7.458 MeV to show that the sequential decay is still dominant but the emitted $\alpha$ have equal energies: a {\it curious result} due to the assumption that all the $\alpha$s are in mutual resonance. In all cases, when the $\alpha$ particles tunnel and are at the outer classical turning point, they are at rest. Thus, technically we have bosons at rest, i.e., a Bose Einstein Condensate, independent if it is a sequential or a direct decay. An experimental verification of the results discussed in this paper would be highly important and the possibility of the existence of the Efimov state (at least as final state interactions) should not be dismissed lightly since it is a phenomenon observed in atomic systems and our 3$\alpha$ cluster system fulfills most of the conditions for the effect to occur. If it does not, then {\it Amen}, but why? Coulomb?

\section*{Acknowledgements}
This work was partially supported by the National Natural Science Foundation of China (No. 11765014, 11605097) and Doctoral Scientific Research Foundation of Inner Mongolia University for the Nationalities (No. BS365, BS400). AB thanks the Chinese Academy of Science, Sinap and Inner Mongolia University for the warm hospitality and support during his stay in China while this work was completed.

%%%%%%%%%%%%%%%%%%%%%%%%%%%%%%%%%%%%%%%%%%%%%%%%%%%%%%%%%%%%

%\section{Acknowledgements}

%---------------------------------------------------------------------------

\end{document}